\documentclass[twocolumn,preprintnumbers,nofootinbib]{revtex4}
\usepackage{amsmath} \usepackage{graphicx} \usepackage{amsfonts}
\usepackage{array} \usepackage{amsthm} \usepackage{bm} 

\usepackage{latexsym}
\usepackage{hyperref}
\hypersetup{colorlinks=false}

\newcommand{\nn}{\nonumber}
\newcommand{\be}{\begin{equation}}
\newcommand{\ee}{\end{equation}}
\newcommand{\ba}{\begin{eqnarray}}
\newcommand{\ea}{\end{eqnarray}}
\newcommand{\bal}{\begin{align}}
\newcommand{\eal}{\end{align}}

\newcommand{\dd}{{\rm d}}
\newcommand{\ii}{{\rm i}}

\newcommand{\om}{\omega}
\newcommand{\al}{\alpha}
\newcommand{\la}{\lambda}

\newcommand{\bt}{\beta}

\newcommand{\ro}{\rho}
\newcommand{\ep}{\epsilon}

\newcommand{\ta}{\theta}

\newcommand{\Si}{\Sigma}
\newcommand{\De}{\Delta}

\newcommand{\Om}{\Omega}

\newcommand{\de}{\delta}

\newcommand{\bw}{\begin{widetext}}
\newcommand{\ew}{\end{widetext}}

\def\abh{black hole }
\def\bh{black holes }
\def\aBH{black hole}
\def\BH{black holes}
\def\RN{Reissner-Nordstr\"om }

\def\Sd{Schwarzschild }

\begin{document}
\begin{flushleft}{\footnotesize Phys. Rev. D \textbf{90}, 064041 (2014)}
\end{flushleft}
\title{Generating rotating regular black hole solutions without complexification}
\author{Mustapha Azreg-A\"inou}
\affiliation {Engineering Faculty, Ba\c{s}kent University,
Ba\u{g}l\i ca Campus, 06810 Ankara, Turkey}


\begin{abstract}
We drop the complexification procedure from the Newman-Janis algorithm and introduce more physical arguments and symmetry properties, and we show how one can generate regular and singular rotating black hole and non-black-hole solutions in Boyer-Lindquist coordinates. We focus on generic rotating regular black holes and show that they are regular on the Kerr-like ring but physical entities are undefined there. We show that rotating regular black holes have much smaller electric charges, and, with increasing charge, they turn into regular non-black-hole solutions well before their Kerr-Newman counterparts become naked singularities. No causality violations occur in the region inside a rotating regular black hole. The separability of the Hamilton-Jacobi equation for neutral particles is also carried out in the generic case, and the innermost boundaries of circular orbits for particles are briefly discussed. Other, but special, properties pertaining to the rotating regular counterpart of the Ay\'on-Beato--Garc\'{\i}a regular static black hole are also investigated.
\end{abstract}


\maketitle

\section{On the Newman-Janis algorithm\label{sec1}}

In this introductory section we comment on two steps in the Newman-Janis algorithm (NJA)~\cite{Janis1}. We first introduce the following general static metric:
\begin{equation}\label{1}
\dd s_{\text{stat}}^2 = G(r)\dd t^2 - \frac{\dd r^2}{F(r)} - H(r)(\dd \ta^2+\sin^2\ta\dd \varphi^2)
\end{equation}

One of the ambiguous steps in the algorithm is complexification of the radial coordinate $r$. This is the step that follows the complex coordinate transformation:
\begin{equation}\label{n1}
r \to r + \ii a \cos \theta,\;u \to  u  - \ii a \cos \theta,
\end{equation}
where ($u,r,\ta,\varphi$) are the advanced null coordinates. Recall that there were already generalizations of this complex coordinate transformation~\cite{ct}, but it seems that the subsequent developments of the NJA and generating methods have not made the matter of further generalizing these transformations a concern. There are as many ways to complexify $r$ as one wants. Here are some examples:
\begin{align}
\label{2}& r^2\to (r + \ii a \cos \theta)(r - \ii a \cos \theta)=r^2+a^2\cos^2 \theta,\\
& \frac{1}{r}\to \frac{1}{2}\Big(\frac{1}{r + \ii a \cos \theta}+\frac{1}{r - \ii a \cos \theta}\Big)=\frac{r}{r^2+a^2\cos^2 \theta},\nn\\
& r^2\to r\sqrt{(r + \ii a \cos \theta)(r - \ii a \cos \theta)}=r\sqrt{r^2+a^2\cos^2 \theta}.\nn
\end{align}
When $a=0$, each right hand side (r.h.s.) reduces to the left hand side (l.h.s) of the same line. Both the first and second types of complexification in~\eqref{2} are used to derive the Kerr solution from the \Sd one: If only one type of complexification is used, the generated rotating solution will not look like the Kerr one. This is the very ambiguity behind nonphysical solutions~\cite{kr} that cannot be written in Boyer-Lindquist coordinates (BLCs) as shown in Ref.~\cite{az2}.

The failure of the last step of the NJA, which consists in bringing the generated rotating solution written in Eddington-Finkelstein coordinates (EFCs) to BLCs by real coordinate transformations, is likely related to the complexification procedure. We have already commented on this point in Ref.~\cite{az2} and have shown that it is not possible, in general, to carry this last step of the NJA. In this work we will raise similar comments and provide another concrete example from the literature~\cite{rrbhs}.

The issue pertaining to complexification has been solved in Ref.~\cite{az3}, where a generic metric formula, not appealing to the complexification procedure, was derived to generate imperfect fluid rotating solutions in BLCs. The metric formula depends on a three-variable function $\Psi(r,\ta,a)$ whose determination depends on the physical problem at hand; that is, it depends on the type of rotating solution one wants to derive. $\Psi$ generally obeys some partial differential equation(s). In the case in which one is generally interested, where the source term in the field equations $T^{\mu\nu}$ is interpreted as an imperfect fluid rotating about a fixed axis, $\Psi$ obeys two linear and nonlinear partial differential equations [see Eqs.~(15) and~(18) of Ref.~\cite{az3} and Eqs.~(4) and~(7) of Ref.~\cite{conf}]. Thus, the essence of our procedure is to reduce the task of determining the rotating counterpart of~\eqref{1} to that of fixing $\Psi$ by solving two partial differential equations. Applications are considered in Refs.~\cite{az3,conf} and in Sec.~\ref{sec3} of this work.

In the following section, we show how one can skip the complexification procedure and we introduce more physical arguments and symmetry properties to derive, based on our previous works~\cite{az3,conf}, rotating metric counterparts of the static ones. We comment again on the last step of the NJA by providing examples from the literature. Application of the rotating metric formula was considered in Refs.~\cite{az3,conf}, where particularly rotating wormholes were derived.

In Sec.~\ref{sec3} we apply the rotating metric formula to derive rotating regular \bh counterparts of static regular ones; then, we discuss their generic physical properties in the first part of Sec.~\ref{sec4}. In the remaining part of the latter section, we specialize to the rotating regular counterpart of the Ay\'on-Beato--Garc\'{\i}a regular static black hole (AGRSBH) and discuss their peculiar physical properties. We conclude in Sec.~\ref{sec5}. Two Appendixes sections have been added: Appendix A was added to check the validity of the Einstein equations and Appendix B to derive the extremality condition for the rotating regular counterpart of the AGRSBH.

\section{Rotating metrics in BLCs \label{sec2}}

Consider the static metric~\eqref{1} to which we partly apply the NJA. After introducing the advanced null coordinates ($u,r,\ta,\varphi$) defined by
\begin{equation*}
    \dd u=\dd t-\dd r/\sqrt{FG},
\end{equation*}
the nonzero components of the resulting inverse metric are of the form $g^{\mu \nu }=l^{\mu} n^{\nu} +  l^{\nu} n^{\mu} -  m^{\mu} {\bar m}^{\nu} -  m^{\nu} {\bar m}^{\mu}$ with
\begin{align}
& l^{\mu } = \delta_r^{\mu}, \nonumber \\
\label{2n}& n^{\mu} = \sqrt{F/G} \,\delta^{\mu}_u - (F/2) \delta^{\mu}_r,  \\
& m^{\mu} = \big( \delta_{\theta}^{\mu} + \frac{ \ii}{\sin \theta}\,\delta_{\varphi}^{\mu} \big)/\sqrt{ 2 H},\nonumber
\end{align}
and $l_{\mu} l^{\mu} = m_{\mu} m^{\mu} = n_{\mu} n^{\mu} = l_{\mu} m^{\mu} =n_{\mu} m^{\mu} =0$
and $l_{\mu} n^{\mu} = - m_{\mu} {\bar m}^{\mu} =1$.

Next, we perform the complex transformation~\eqref{n1} by which $\delta_{\nu}^{\mu}$ transform as vectors:
\begin{equation}\label{tr1}
\delta_r^{\mu}\to\delta_r^{\mu},\, \delta_u^{\mu}\to\delta_u^{\mu},\, \delta_{\theta}^{\mu}\to\delta_{\theta}^{\mu}+\ii a\sin\ta(\delta_u^{\mu}-\delta_r^{\mu}),\, \delta_{\varphi}^{\mu}\to\delta_{\varphi}^{\mu},
\end{equation}
and we assume that \{$G,F,H$\} transform to \{$A,B,\Psi$\}:
\begin{equation}\label{tr2}
\{G(r),F(r),H(r)\}\to \{A(r,\ta,a),B(r,\ta,a),\Psi(r,\ta,a)\},
\end{equation}
where \{$A,B,\Psi$\} are three-variable real functions, to be fixed later. For the purpose of this paper we subject them to the following constraints:
\begin{multline}\label{n3}
\lim_{a\to 0}A(r,\ta,a)=G(r),\;\;\lim_{a\to 0}B(r,\ta,a)=F(r),\\
\lim_{a\to 0}\Psi(r,\ta,a)=H(r),
\end{multline}
to recover~\eqref{1} in the limit $a\to 0$. For other purposes, see Refs.~\cite{conf,az3}. We thus depart from the usual NJA, which fixes the expressions of \{$A,B,\Psi$\} by complexification of the radial coordinate $r$. In our procedure, \{$A,B,\Psi$\} will be fixed using other criteria and physical arguments.

The effect of the transformation~\eqref{n1} on ($l^{\mu },n^{\mu},m^{\mu}$) is the composition of the transformations~\eqref{tr1} and~\eqref{tr2} on $\delta_{\nu}^{\mu}$ and on \{$G(r),F(r),H(r)$\}, respectively:
\begin{align}
& l^{\mu } = \delta_r^{\mu}, \nonumber \\
\label{4n}& n^{\mu} = \sqrt{B/A} \,\delta^{\mu}_u - (B/2) \delta^{\mu}_r,  \\
& m^{\mu} = \big[ \delta_{\theta}^{\mu}+\ii a\sin\ta(\delta_u^{\mu}-\delta_r^{\mu}) + \frac{ \ii}{\sin \theta}\,\delta_{\varphi}^{\mu} \big]/\sqrt{ 2 \Psi}.\nonumber
\end{align}
This yields the transformed inverse metric
\begin{align}
& g^{u u }(r,\ta) = -\frac{a^2 \sin^2\theta}{ \Psi}, \;\;
g^{u \varphi}(r,\ta) = -\frac{a}{ \Psi} , \nonumber  \\
& g^{\varphi \varphi}(r,\ta) = -\frac{1}{\Psi\sin^2 \theta}  , \;\;
g^{\theta \theta}(r,\ta) = -\frac{1}{ \Psi},  \nonumber \\
& g^{rr}(r,\ta) = -B-\frac{a^2 \sin ^2 \theta }{ \Psi} , \;\;
g^{r \varphi}(r,\ta) = \frac{a}{\Psi} ,\nonumber  \\
& g^{u r }(r,\ta) = \sqrt{\frac{B}{A}}+\frac{a^2 \sin ^2\theta }{\Psi},
\end{align}
and then the rotating metric in EFCs
\begin{multline}\label{n5}
\dd s^2 = A \dd u^2+2 \frac{\sqrt{A}}{\sqrt{B}} \dd u \dd r+2 a\sin ^2\theta  \Big(\frac{\sqrt{A}}{\sqrt{B}}-A\Big) \dd u \dd \varphi \\- 2 a\sin ^2\theta  \frac{\sqrt{A}}{\sqrt{B}} \dd r \dd \varphi-\Psi  \dd \theta^2 \\- \sin ^2\theta  \Big[\Psi +a^2\sin ^2\theta  \Big(2 \frac{\sqrt{A}}{\sqrt{B}}-A\Big)\Big] \dd \varphi^2.
\end{multline}

The final but crucial step is to bring~\eqref{n5} to BLCs by a global coordinate transformation that is usually taken of the form
\begin{equation}\label{n6}
\dd u=\dd t + \la(r)\dd r,\,\dd\varphi=\dd\phi +\chi(r)\dd r,
\end{equation}
where \{$\la,\chi$\} must depend on $r$ only to ensure integrability of~\eqref{n6}: It is easy to check that, in this, case one can integrate the two equations to obtain global coordinates $u(t,r)$ and $\varphi(\phi,r)$. As explained in Sec.~\ref{sec1}, the usual NJA fails, in general, to bring~\eqref{n5} to BLCs since in the NJA, \{$A,B,\Psi$\} are fixed by the complexification of $r$ and there remain no free parameters or functions to act on to achieve the transformation to BLCs.

This is no longer the case in our procedure since \{$A,B,\Psi$\} are still unknown functions and we can achieve the transformation to BLCs. This is indeed the case; taking
\begin{equation}\label{n7}
\la(r)=- \frac{(K+a^2)}{FH+a^2},\,\chi(r)=-\frac{a}{FH+a^2},
\end{equation}
where
\begin{equation}\label{n8}
K(r)\equiv \sqrt{\frac{F(r)}{G(r)}}H(r),
\end{equation}
the metric~\eqref{n5} is brought to BLCs
provided we choose
\begin{equation}\label{9}
A(r,\ta)=\frac{(FH+a^2\cos^2\ta) \Psi }{(K+a^2 \cos^2\ta)^2},\, B(r,\ta)=\frac{FH+a^2\cos^2\ta}{\Psi}.
\end{equation}
Finally, the desired form of the rotating solution in BLCs is~\cite{az3,conf}
\begin{align}
&\dd s^2 = \frac{(F H+a^2 \cos ^2\theta ) \Psi \dd t^2}{(K+a^2 \cos ^2\theta )^2}
    -\frac{\Psi \dd r^2}{F H+a^2}\nn\\
&+2 a \sin ^2\theta\Big[\frac{K- FH}{(K+a^2 \cos ^2\theta )^2}\Big]\Psi\dd t\dd \phi -\Psi\dd \ta^2\nonumber\\
\label{10}&-\Psi\sin ^2\theta\Big[1+a^2\sin ^2\theta\frac{2K- FH
    +a^2\cos ^2\theta }{(K+a^2\cos ^2\theta )^2}\Big]\dd \phi^2.
\end{align}
This latter metric is brought to Kerr-like forms~\cite{az3,conf}
\begin{align}
&\dd s^2 =\frac{\Psi}{\ro^2}\Big[\Big(1-\frac{2f}{\ro^2}\Big)\dd t^2-\frac{\ro^2}{\De}\,\dd r^2\nn\\
\label{10a}&+\frac{4af \sin ^2\theta}{\ro^2}\,\dd t\dd \phi-\ro^2\dd \ta^2-\frac{\Si\sin ^2\theta}{\ro^2}\,\dd \phi^2\Big]
\end{align}
\begin{align}
\label{10b}&\dd s^2 =\frac{\Psi}{\ro^2}\Big[\frac{\De}{\ro^2}(\dd t-a\sin^2\ta\dd \phi)^2-\frac{\ro^2}{\De}\,\dd r^2-\ro^2\dd \ta^2\nn\\
&-\frac{\sin^2\ta}{\ro^2}\,[a\dd t-(K+a^2)\dd \phi]^2\Big].
\end{align}
on performing the following variable changes:
\begin{align}
&\ro^2\equiv K+a^2 \cos^2\ta,\;\;2f(r)\equiv K-FH\nn\\
\label{10c}&\De(r)\equiv FH+a^2,\;\;\Si\equiv (K+a^2)^2-a^2\De\sin^2\ta.
\end{align}

In Eqs~\eqref{10}--\eqref{10b} $\Psi(r,\ta,a)$ remains an unknown function. If the source term $T^{\mu\nu}$ is interpreted as an imperfect fluid rotating about the $z$ axis, $\Psi$ obeys the two nonlinear and linear partial differential equations (15) and (18) of Ref.~\cite{az3} to which some particular solutions were found in Refs.~\cite{az3,conf}. These equations take the following forms:
\begin{align}
\label{n9}&(K+a^2 y^2)^2 (3\Psi_{,r}\Psi_{,y^2} -2\Psi \Psi_{,ry^2}) =3a^2K_{,r}\,\Psi^2,\\
&[{K_{,r}}^2+K(2-K_{,rr})-a^2y^2(2+K_{,rr})]\Psi\nn\\
\label{n10}&+(K+a^2y^2)(4y^2\Psi_{,y^2}-K_{,r}\Psi_{,r})=0,
\end{align}
where the indexical notation for derivatives $\Psi_{,ry^2}\equiv \partial^2\Psi/\partial r\partial y^2$, $K_{,r}\equiv \partial K/\partial r$, etc., has been used and $y\equiv \cos\ta$.

The nonlinear differential equation~\eqref{n9} is just $G_{r\ta}=0$, where $G_{\mu\nu}$ is the Einstein tensor, and the linear differential equation~\eqref{n10} ensures consistency of the field equations $G_{\mu\nu}=T_{\mu\nu}$ with the expression of $T^{\mu\nu}$:
\begin{equation}\label{rv1}
T^{\mu\nu}=\ep e^{\mu}_te^{\nu}_t+p_re^{\mu}_re^{\nu}_r+p_{\ta}e^{\mu}_{\ta}e^{\nu}_{\ta}+p_{\phi}e^{\mu}_{\phi}e^{\nu}_{\phi},
\end{equation}
where $e^{\mu}_t$ is the four-velocity vector of the fluid, $\ep$ is the density, ($p_r,\,p_{\ta},\,p_{\phi}$) are the components of the pressure, and the basis $(e_t,\,e_r,\,e_{\ta},\,e_{\phi})$ is dual to the 1-forms defined in~\eqref{10b} [see Eqs.~(16) and~(17) of Ref.~\cite{az3}]:
\begin{align}
&\om^t\equiv \sqrt{\Psi\De}(\dd t-a\sin^2\ta\dd \phi)/\ro^2, \nn\\
\label{rv2}&\om^r\equiv -\sqrt{\Psi}\dd r/\sqrt{\De},\,
\om^{\ta}\equiv -\sqrt{\Psi}\dd \ta, \\
&\om^{\phi}\equiv -\sqrt{\Psi}\sin\ta [a\dd t-(K+a^2)\dd \phi]/\ro^2. \nn
\end{align}

We once more comment on the NJA by providing an explicit example from the literature where it is not possible to carry out the last step that consists in bringing the rotating metric from EFCs to BLCs.

In Eqs.~(20) of Ref.~\cite{rrbhs}, each r.h.s. is a total differential (exact differential), provided the functions $\la$ and $\chi$ depend only on $r$ as in~\eqref{n6}. Unfortunately, this is not the case in the final expressions of $\la$ and $\chi$ given in the r.h.s.'s of Eqs.~(21) of Ref.~\cite{rrbhs}, which generally depend on both ($r,\ta$): Only in the trivial case $Q=0$, which corresponds to the \Sd solution, do $\la$ and $\chi$ depend only on $r$.

If $\la$ and $\chi$ depend on both ($r,\ta$) then $\partial \la/\partial \ta \neq 0$ and $\partial \chi/\partial \ta \neq0$, so the conditions of integrability are no longer satisfied and it is not possible to integrate Eqs.~(20) of Ref.~\cite{rrbhs} to obtain global coordinates $u(t,r,\ta)$ and $\phi(\phi,r,\ta)$. Consequently, if $Q\neq 0$, the set of Eqs.~(20) of Ref.~\cite{rrbhs} does not constitute a global coordinate transformation and the final metric Eqs.~(22) of Ref.~\cite{rrbhs} is not equivalent to the metric~(19) of Ref.~\cite{rrbhs}, which is given in EFCs. Said otherwise, if $Q\neq 0$, it is not possible, by a global coordinate transformation, to bring metric~(19) in EFCs to a rotating metric in BLCs.

Other examples from the literature of such a failure to carry out the last step of the NJA, that is, examples where the EFC-to-BLC transformation has been carried out by noncoordinate transformations, as in Ref.~\cite{rrbhs}, are found in Refs.~\cite{kr,nonc,rap} and are certainly due to the type(s) of complexification used. This is a general drawback of the NJA since it does not fix \textit{a priori} the type(s) of complexification needed to carry the EFC-to-BLC transformation.

Such noncoordinate transformations used by some authors~\cite{kr,rrbhs,nonc,rap} to carry the EFC-to-BLC transformation of the NJA could, however, be seen as an added trick to the NJA, which is by itself a trick to obtain rotating solutions from static ones. But, this may lead to nonphysical solutions as in Ref.~\cite{kr} or to modified theories of general relativity, that is, to solutions with a ``set of field equations \ldots different from the Einstein equations \ldots" as in~\cite{rap}.

\section{Rotating regular black holes\label{sec3}}

To our knowledge, all regular \bh in classical general relativity have $G=F$ and $H=r^2$~\cite{IrinaD}-\cite{reg}. In the case $G=F$, a general prescription for determining imperfect fluid rotating (about the $z$ axis) regular \bh is given in Sec.~3 of Ref.~\cite{az3}; we outline it here.

Equation~\eqref{n8} implies $K=H=r^2$. Now, it is easy to check that
\begin{equation}\label{s2}
    \Psi=r^2+a^2\cos^2\ta
\end{equation}
is one of the solutions to~\eqref{n9} and~\eqref{n10} satisfying~\eqref{n3} with $A$ and $B$ given by~\eqref{9}. Using~\eqref{10c}, with $K=r^2$, along with~\eqref{s2} in~\eqref{10a}, the regular rotating counterpart \abh of a regular static one ($G=F$ and $H=r^2$) takes the compact form
\begin{align}
&\dd s^2 =\Big(1-\frac{2f}{\ro^2}\Big)\dd t^2-\frac{\ro^2}{\De}\,\dd r^2\nn\\
\label{r1}&+\frac{4af \sin ^2\theta}{\ro^2}\,\dd t\dd \phi-\ro^2\dd \ta^2-\frac{\Si\sin ^2\theta}{\ro^2}\,\dd \phi^2\\
&\ro^2=r^2+a^2\cos^2\ta,\;\;2f=r^2(1-F)\nn\\
&\De = r^2F+a^2=r^2-2f+a^2\nn\\
&\Si = (r^2+a^2)^2-a^2\De\sin^2\ta .\nn
\end{align}

In Appendix A, we show that the rotating solution~\eqref{r1} satisfies Einstein equations $G_{\mu\nu}=T_{\mu\nu}$, where $T^{\mu\nu}$ is of the form~\eqref{rv1}.

In this paper, we will discuss the general solution~\eqref{r1} as well as the regular rotating counterpart of the AGRSBH~\cite{reg}. In our notation, the AGRSBH, which was derived in Ref.~\cite{reg}, takes the form
\begin{equation}\label{s1}
    G=F=1-\frac{2Mr^2}{(r^2+Q^2)^{3/2}}+\frac{Q^2r^2}{(r^2+Q^2)^2},\;\;H=r^2.
\end{equation}
Its regular rotating counterpart, given by~\eqref{r1} and~\eqref{s1}, reduces to the Kerr metric if $Q=0$, where, in this case, $2f=2Mr$, $\De =r^2-2Mr+a^2$, and $\Si =(r^2+a^2)\ro^2+2Ma^2r\sin^2\ta$.

\section{Physical properties\label{sec4}}

In this section we discuss the general properties of the regular rotating solution~\eqref{r1} for any regular static black hole $F$ as well as its special properties for the AGRSBH, where $F$ is given by~\eqref{s1}.

Since metric~\eqref{r1} generates all types of rotating solutions, general properties of singular rotating \bh are also investigated. However, we focus mostly on rotating regular \BH. The first part of this section is devoted to a general discussion, and the second one is concerned with the rotating regular counterpart of the AGRSBH.

\subsection{General physical properties}

Notice that the only difference between Kerr's metric and ~\eqref{r1} resides in the values of ($f,\De,\Si$). Moreover, and this is an important point in our method, metric~\eqref{r1} is a fresh formula; that is, it applies to all static (being regular or not) \bh of the form~\eqref{1} with $G=F$ and $H=r^2$, and the only task one has to perform is to evaluate $2f=r^2(1-F)$, $\De = r^2F+a^2$, and $\Si = (r^2+a^2)^2-a^2\De\sin^2\ta$, knowing $F$. There are no notions of complexification associated with the different forms of $F$, while the application of the NJA necessitates different ways of complexification for each different form of $F$ and the final rotating metric may only be given in EFCs because of the nonexistence of coordinate transformations bringing it to BLCs, as were the cases in Refs.~\cite{kr,rrbhs,nonc,rap}.\\

\paragraph*{\textbf{1. Scalar invariants and stress-energy tensor.}} We keep on doing generalities which apply to all rotating regular \bh of the form~\eqref{r1} (other conclusions apply also to singular solutions). Static regular \bh have de Sitter--like behavior near $r=0$~\cite{IrinaD}-\cite{reg}:
\begin{equation}\label{e1}
    F\sim 1-Cr^2\text{ and }C>0\; (r\to 0),
\end{equation}
which results in
\begin{equation}\label{e2}
    f\sim Cr^4\; (r\to 0).
\end{equation}

The curvature scalar $R$ of the regular rotating solution~\eqref{r1} reads
\begin{equation}\label{e3}
    R=\frac{2(1-F)-4rF^{\,\prime}-r^2F^{\,\prime\prime}}{\rho ^2},
\end{equation}
(here, $F^{\,\prime}\equiv \dd F/\dd r$ etc.), which is manifestly regular off the ring $\ro^2=0$. Following a procedure used in Refs.~\cite{conf,az3}, it is easy to show that $R$ is also regular on the ring $\ro^2=0$. In fact, let $\mathcal{C}$ be any path in the $yr$ plane (the $y=\cos\ta$ axis is horizontal, and the $r$ axis is vertical) through the ring $\ro^2=0$; that is, $\mathcal{C}$: $r=ah(y)$ and $h(0)=0$. Then, using~\eqref{e1}, we obtain
\begin{equation}\label{e4}
    \lim_{y\to0}R=\frac{12Ch_{,y}(0)^2}{h_{,y}(0)^2+1}=\frac{12C}{1+g_0{}^2},
\end{equation}
where $g_0\equiv 1/h_{,y}(0)$. Thus, whatever the value of the slope of $\mathcal{C}$ at $y=0$, $h_{,y}(0)$, the value of the limit $\lim_{y\to0}R$ is finite. Since the limit depends on the value of $h_{,y}(0)$, $R$ is undefined; however, it is finite and regular on the ring $\ro^2=0$. See case (1) of Ref.~\cite{az3} for a more general discussion.

The mathematical expression of the Kretschmann scalar $K\equiv R_{\al\bt\mu\nu}R^{\al\bt\mu\nu}$ is sizable, so we will not provide here. This scalar is regular everywhere, including the ring $\ro^2=0$. On any path $\mathcal{C}$: $r=ah(y)$ and $h(0)=0$ through the ring $\ro^2=0$, we have
\begin{align}
\lim_{y\to0}K=&\frac{24 C^2 h_{,y}(0)^4}{[1+h_{,y}(0)^2]^6} [6-2 h_{,y}(0)^2+11 h_{,y}(0)^4\nn\\
\label{e5}\quad &+4 h_{,y}(0)^6+h_{,y}(0)^8]\\
\label{e6}\quad = &\frac{24 C^2[1+4 g_0{}^2+11 g_0{}^4-2 g_0{}^6+6g_0{}^8]}{[1+g_0{}^2]^6}.
\end{align}
This limit is finite for all paths $\mathcal{C}$: $r=ah(y)$ and $h(0)=0$ through the ring $\ro^2=0$ but it remains undefined, for it depends on $h_{,y}(0)$.

The components ($\ep,p_r,\,p_{\ta},\,p_{\phi}$) of the stress-energy tensor (SET) $T^{\mu\nu}$ are given by Eqs.~(13) and~(14) of Ref.~\cite{conf}, taking $p^2=0$ (these have been evaluated in Refs.~\cite{Gurses,Magli}, too):
\begin{equation}\label{11ab}
\ep =-p_r=\frac{2(rf^{\,\prime} -f)}{\rho ^4},\quad p_{\ta}=p_{\phi}=-p_r-\frac{f^{\,\prime\prime}}{\rho ^2},
\end{equation}
(here, $f^{\,\prime}\equiv \dd f/\dd r$), which, despite their appearance, have been shown not to diverge on the ring $\ro^2=0$ because of de Sitter--like behavior~\eqref{e1} and~\eqref{e2} near $r=0$ of the static regular \bh [see paragraph following Eq.~(14) of Ref.~\cite{conf} and case (1) of Ref.~\cite{az3} that use the same procedure as the one by which~\eqref{e4}, \eqref{e5} and~\eqref{e6} were derived]. These components are, however, undefined on the ring $\ro^2=0$, in that the limits $\lim_{r\to0,\ta\to\pi/2}(\ep,p_r,\,p_{\ta},\,p_{\phi})$ do not exist.

Because of the relation $p_r=-\ep$, these solutions can also be used as regular cores to match other rotating external solutions~\cite{conf}. Note that the NJA was first devised to generate exterior rotating solutions but later was applied to generate rotating interior metrics which were matched to the exterior Kerr one~\cite{Herrera,Via}.

Notice from~\eqref{11ab} that, since $f$ does not depend on the rotation parameter $a$, $\ep$ has the same sign as its static counterpart $\ep_{\text{stat}}$: $\ep=(r^4/\rho ^4)\ep_{\text{stat}}$. This remark is very relevant for the determination of the energy conditions of rotating regular \BH. For the rotating regular \abh solution~\eqref{r1} with $F$ given by~\eqref{s1}, it was reported that its static counterpart \abh satisfies the weak energy condition~\cite{KB}, that is, $\ep_{\text{stat}}\geq 0$; we thus conclude that $\ep\geq 0$. Because of de Sitter--like behavior near $r=0$ of the static regular \aBH, this latter conclusion is valid for all rotating regular \bh near $r=0$ where $rf^{\,\prime} -f\simeq 3 Cr^4$.

It is straightforward to check that the components of the SET given by~\eqref{11ab} approach those of the Kerr-Newman \abh in the limit $r\to\infty$ if $F$ approaches the \RN limit.

The function $f^{\,\prime\prime}$ is 0 only for Reissner-Nordstr\"om-like solutions of the form $F=1+c_1/r+c_2/r^2$. For all other regular or singular solutions $f^{\,\prime\prime}\neq 0$ and, by~\eqref{11ab}, $p_{\ta}=p_{\phi}\neq \ep =-p_r$, so the fluid is never perfect.\\

\paragraph*{\textbf{2. Static limit: Horizons.}} The mass of the rotating solution, being regular or not, is that of the static one. This is obvious from~\eqref{r1} for if $F$ admits a Taylor expansion of the form $F=1-2M/r+\cdots$ at spatial infinity, then the two metric functions $g_{tt}=$ and $1/g_{rr}$ of the rotating solution~\eqref{r1} admit the same expansion as $r\to \infty$.

The static limit, which is the 2-surface on which the timelike Killing vector $t^{\mu}=(1,0,0,0)$ becomes null, corresponds to $g_{tt}(r_{\text{st}},\ta)=0$ leading to $2f=\ro^2$ or simply the following separable equation
\begin{equation}\label{sl1}
a^2\cos^2\ta =-r_{\text{st}}{}^2F(r_{\text{st}})
\end{equation}
as in the Kerr and Kerr-Newman cases. Observers can remain static only for $r>r_{\text{st}}(\ta)$.

The event horizon $r_+$, which sets a limit for stationary observers, and the inner apparent one $r_-$ are solutions to $g^{rr}(r_{\pm})=0$ implying $\De (r_{\pm})=0$:
\begin{equation}\label{sl2}
    r_{\pm}{}^2F(r_{\pm})+a^2=0.
\end{equation}
It is clear from these last two equations that the static limit and event horizon intersect only at the two poles $\ta=0$ and $\ta=\pi$, where $r_{\text{st}}=r_+$, as in Kerr and Kerr-Newman solutions. The resolution of~\eqref{sl2} provides $r_{\pm}$ as functions of the charges ($M,Q,\cdots$), on which $F$ depends and of $a^2$ only, so that $r_{\pm}$ do not depend on $\ta$.

\begin{figure}[!htb]
\centering
  \includegraphics[width=0.45\textwidth]{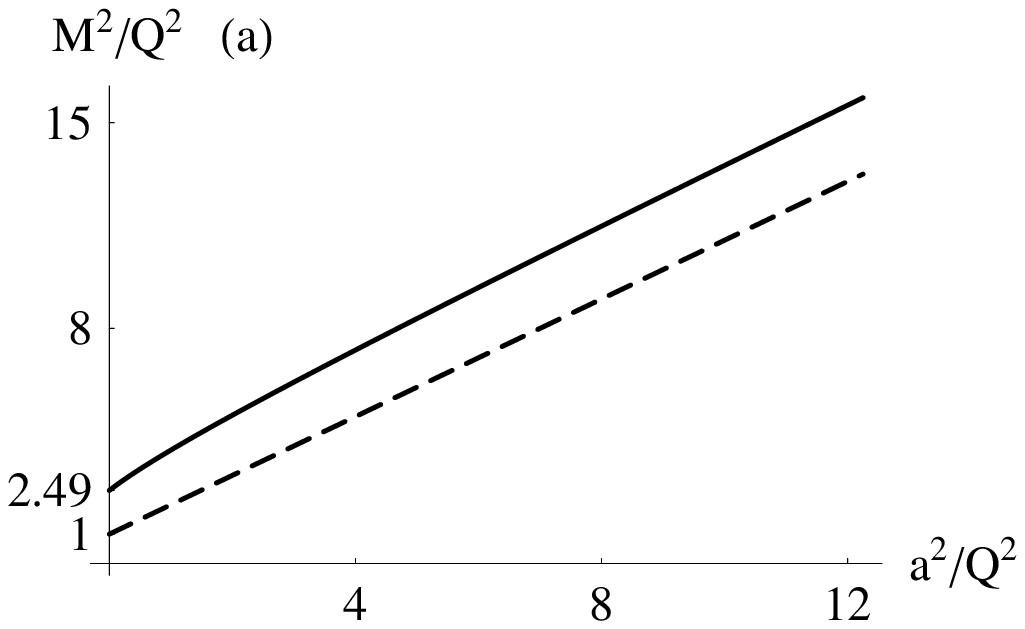} \includegraphics[width=0.4\textwidth]{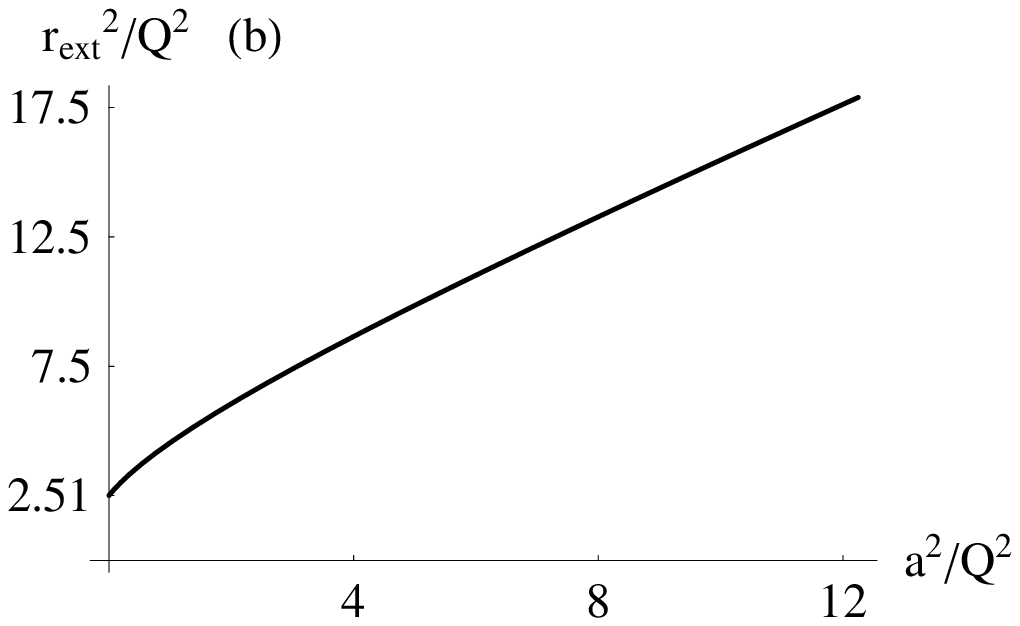}
  \caption{\footnotesize{(a) Using different horizontal and vertical scales, we show in the ($a^2,M^2$) plane the extremality condition. Continuous plot: Rotating regular black hole~\eqref{r1} with $F$ given by~\eqref{s1}. The \abh region is above this curved line. The curve itself represents an extremal \abh and the region beneath it represents regular non-black-hole solutions. Dashed plot: Rotating Kerr-Newman \aBH. The Kerr-Newman \abh region is above this straight line of the equation $M^2/Q^2=a^2/Q^2+1$. Notice that the region between the two plots corresponds to $Q^2<M^2$ which is within the \abh region for the Kerr-Newman solution but within the non-black-hole region ($\forall\, a^2\geq 0$) for the rotating regular black hole~\eqref{r1}. This is a parametric plot of $1/(2s)^2$ versus $u^2$ (see Appendix B). (b) The common radius $r_{\text{ext}}{}^2$ of the merging horizons vs. $a^2$. For $a^2=0$, we have $r_{\text{ext}}{}^2/Q^2\simeq 2.51155$ yielding $r_{\text{ext}}/|Q|\simeq 1.58$ as found in~\cite{reg}. This is a parametric plot of $t-1$ versus $u^2$ (see Appendix B).}}\label{Fig1}
\end{figure}

It is well known that if $Q^2<M^2$, a Kerr-Newman solution may have the properties of a rotating \aBH; this happens if $0<a^2\leq M^2-Q^2$; otherwise ($a^2> M^2-Q^2$), the solution is a naked singularity. In the case where $Q^2\geq M^2$, a Kerr-Newman solution is always a naked singularity for all $a^2>0$. As we shall see in the next section, even in the case where $Q^2<M^2$, it is possible to have no rotating regular \bh for all $a^2$ but only regular non-black-hole solutions given by~\eqref{r1}, as is the case shown in Fig.~\ref{Fig1} (a), which is a plot of the extremality condition in the ($a^2/Q^2,M^2/Q^2$) plane. A similar conclusion was made in Ref.~\cite{KB}. If the function $\mathcal{F}(r)\equiv r^2F\,(=\De -a^2)$, which is 0 at $r=0$ for a static regular \abh (respectively, constant for a singular \aBH) and $\mathcal{F}\to\infty$ as $r\to\infty$, has some negative minimum value $\mathcal{F}_{\text{min}}$ on the range of $r$, then there is always a \abh solution if
\begin{equation}
    0<a^2\leq |\mathcal{F}_{\text{min}}|
\end{equation}
and a non-black-hole solution (respectively, a naked singularity) for
\begin{equation}
a^2> |\mathcal{F}_{\text{min}}|.
\end{equation}
The extremality condition is
\begin{equation}
a^2= |\mathcal{F}_{\text{min}}|
\end{equation}
which provides a relation between the charges ($M,Q,\cdots$) and $a^2$.

\begin{figure}[!htb]
\centering
  \includegraphics[width=0.4\textwidth]{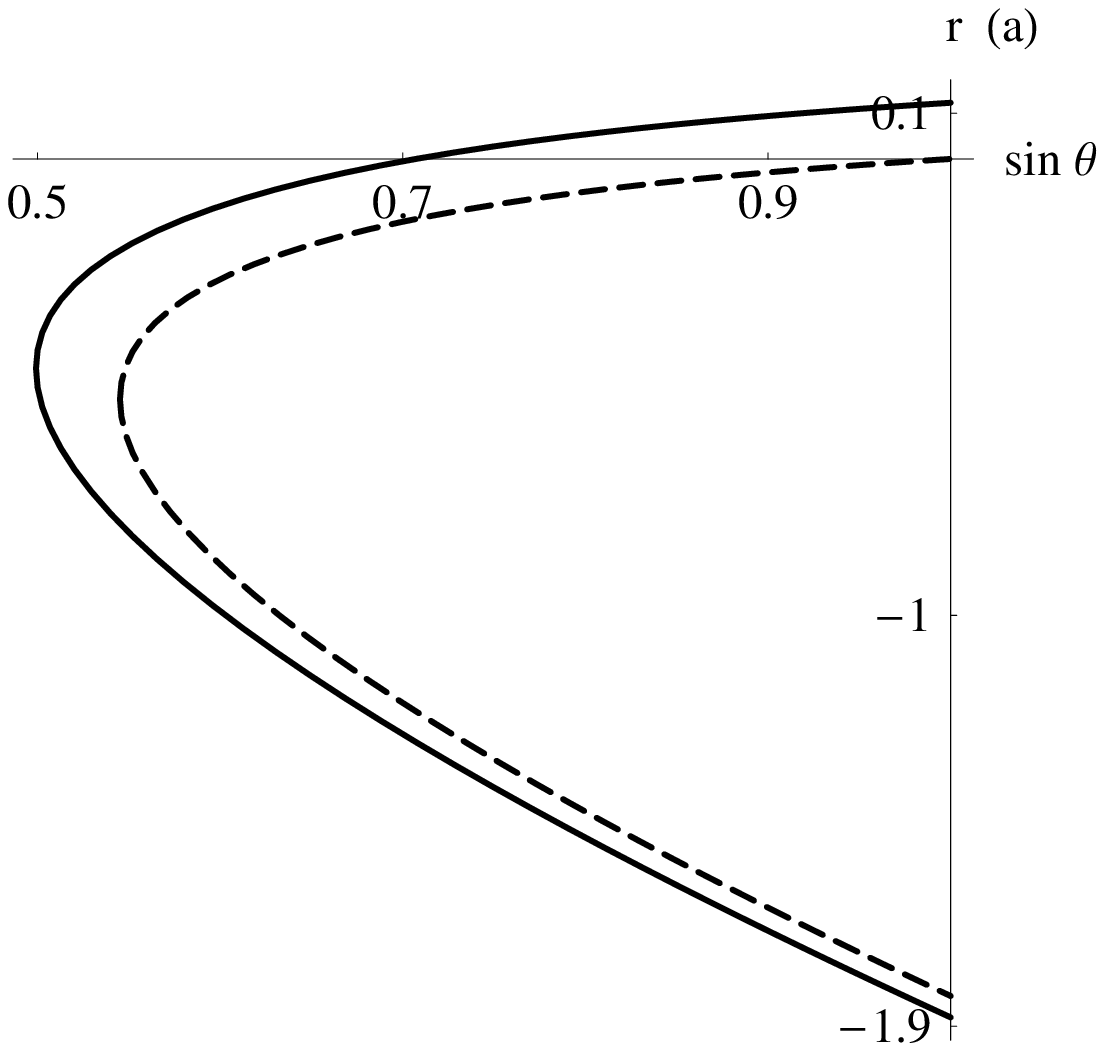} \includegraphics[width=0.4\textwidth]{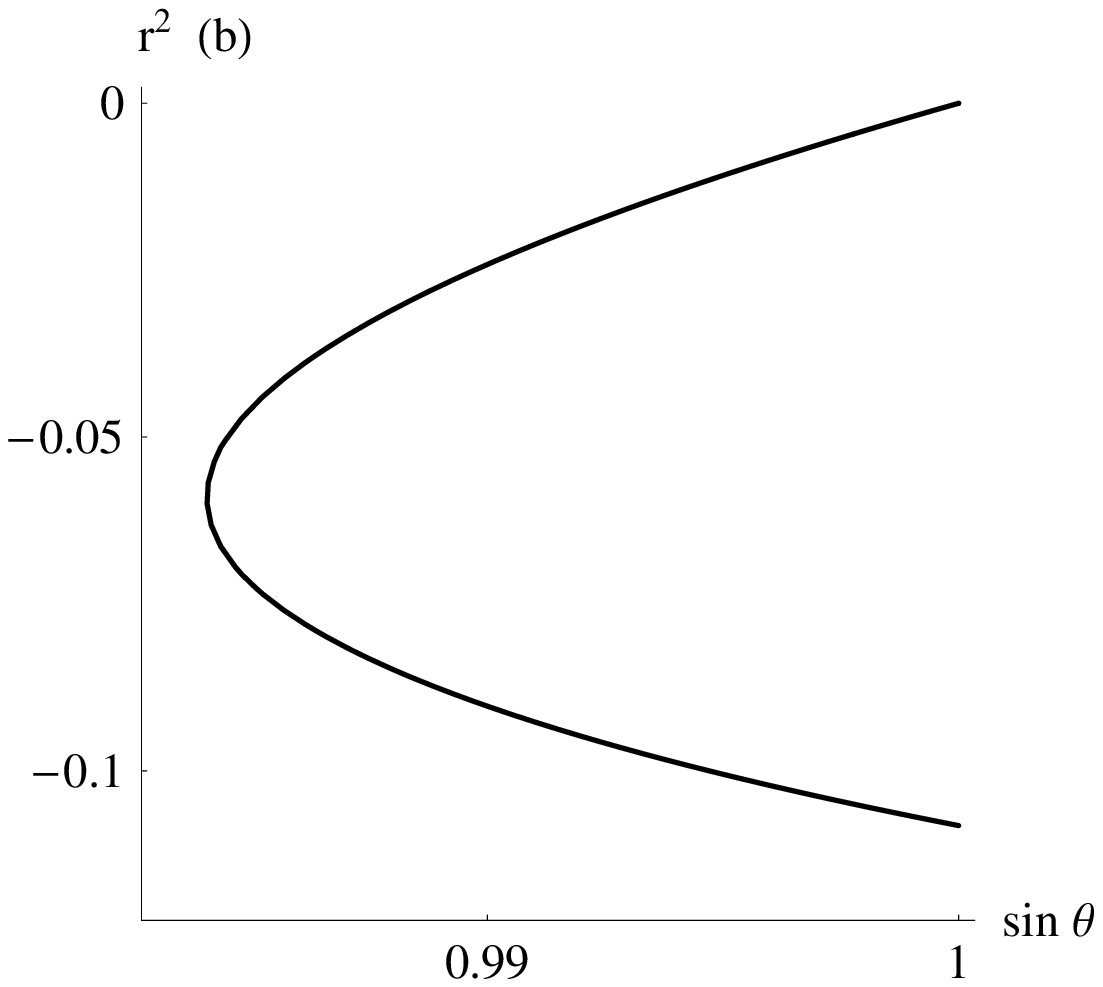}
  \caption{\footnotesize{For all the plots we took $M^2=16$, $Q^2=1$, and $a^2=1$ corresponding, according to Fig.~\ref{Fig1} (a), to the \abh region for Kerr, Kerr-Newman, and the rotating regular \abh solution~\eqref{r1} with $F$ given by~\eqref{s1}. (a) Implicit plot of $\Si(\sin\ta,r)=0$, where $\Si = (r^2+a^2)^2-a^2\De\sin^2\ta$ and $\De = r^2F+a^2$, for Kerr \abh (dashed plot: $F=1-2M/r$), Kerr-Newman \abh (continuous plot: $F=1-2M/r+Q^2/r^2$), and the rotating regular \abh solution~\eqref{r1} with $F$ given by~\eqref{s1} (the point $\sin\ta =1$ and $r=0$). Causality violations and CTCs occur on the right of each curve where $g_{\phi\phi}>0$. The Kerr \abh has CTCs for $r<0$ only while the Kerr-Newman one has CTCs for both signs of $r$.  For the rotating regular \abh solution~\eqref{r1} with $F$ given by~\eqref{s1} no causality violations or CTCs occur since $g_{\phi\phi}<0$ [except at the point ($\sin\ta =1$ and $r=0$) where $\Si=0$ and $g_{\phi\phi}$ is undefined]. (b) Implicit plot of $\Si(\sin\ta,r^2)=0$ for the rotating regular \abh solution~\eqref{r1} with $F$ given by~\eqref{s1}. The plot confirms that solutions to $\Si(\sin\ta,r^2)=0$ for $\sin\ta \neq 1$ are such that $r^2<0$.}}\label{Fig2}
\end{figure}

\paragraph*{\textbf{3. Causality issues.}} It is also well known that causality violations occur in Kerr and Kerr-Newman \BH, as depicted in Fig.~\ref{Fig2} (a). Causality violations and closed timelike curves (CTCs) are possible if, in~\eqref{r1}, $g_{\phi\phi}=-\Si\sin ^2\theta/\ro^2>0$. Since $\sin ^2\theta/\ro^2$ is not negative, for simplicity, we investigate the sign of $\Si =(r^2+a^2)^2-a^2(r^2F+a^2)\sin^2\ta$. Figure~\ref{Fig2} (a) is a plot of $r$ versus $\sin\ta$, where, for a given $\ta$, $r$ is a solution to $\Si(\sin\ta,r)=0$, and Fig.~\ref{Fig2} (b) is a plot of $r^2$ versus $\sin\ta$, where $r^2$ is a solution to $\Si(\sin\ta,r^2)=0$. Causality violations occur on the right of each plot in Fig.~\ref{Fig2} (a), where the dashed curve corresponds to the Kerr \abh and the continuous one corresponds to the Kerr-Newman \aBH, for which CTCs exist even for $r>0$ [in contrast with the Kerr hole, where CTCs are possible for $r<0$ only, as depicted in Fig.~\ref{Fig2} (a)]. In Fig.~\ref{Fig2} (a), the plot of $\Si=0$ for the rotating regular hole~\eqref{r1} where $F$ is given by~\eqref{s1} (the rotating regular counterpart of the AGRSBH) is the point $\sin\ta =1$ and $r=0$. Since for $\sin\ta=0$, $\Si>0$, this implies that $\Si\geq 0$, at least for the values of the parameters we have chosen $M^2=16$, $Q^2=1$, and $a^2=1$, corresponding, according to Fig.~\ref{Fig1} (a), to the \abh region for the rotating regular counterpart of the AGRSBH. This shows that there are no causality violations for this \abh since the sign of $g_{\phi\phi}$ cannot go positive, that is, the Killing vector $\phi^{\mu}=(0,0,0,1)$, of norm $g_{\phi\phi}$, cannot become timelike.

Let us see under which general conditions the above conclusion remains valid. Notice that causality violations are not expected in the region $r>r_+$ or in in the region between the horizons, since there, $\De<0$, yielding $\Si>0$ and $g_{\phi\phi}<0$. Let $r<r_-$. The condition $\Si>0$ yields $(r^2+a^2)^2>a^2(r^2F+a^2)\sin^2\ta$. Since $\De =r^2F+a^2>0$ for $r<r_-$, if we can show that
\begin{equation}\label{c1}
(r^2+a^2)^2>a^2(r^2F+a^2),
\end{equation}
this results in $\Si>0$. Simplifying~\eqref{c1}, we bring it to
\begin{equation}\label{c2}
r^2-a^2F(r)+2a^2>0.
\end{equation}
The condition~\eqref{c2} is satisfied at $r=0$ and $r=r_-$, where its l.h.s. is $a^2$ and $r_-{}^2+(a^4/r_-{}^2)+2a^2$, respectively. Here, we have used $F(0)=1$ and $\De (r_-) =r_-{}^2F(r_-)+a^2=0$. Thus, if $r=\ep a$ or $r=r_--\eta$, where $\ep$ is a small positive or negative number\footnote{This same result could be achieved setting $r=\ep a$ and $\ta=(\pi/2)+\de$, where $\de$ is a small positive or negative number, yielding $\Si\simeq a^4(\ep^2+\de^2)$.} and $\eta$ is a small positive number, there are no causality violations for all rotating regular \BH.

It might be true that the condition~\eqref{c2} holds for all $r<r_-$, including negative values down to $-r_-$. The derivative of the l.h.s. of~\eqref{c2} is
\begin{equation}\label{c3}
    2r-a^2F^{\,\prime},
\end{equation}
which vanishes at $r=0$. Because of the de Sitter behavior~\eqref{e1}, the function $F$ approaches 1 from below, resulting in $F^{\,\prime}<0$ near the origin. If $F^{\,\prime}<0$ holds for all $0<r<r_-$, then $ 2r-a^2F^{\,\prime}>0$ and the l.h.s. of~\eqref{c2} increases from $a^2$ to $r_-{}^2+(a^4/r_-{}^2)+2a^2$; hence, no causality violations occur for $0<r<r_-$. Even if $F^{\,\prime}<0$ fails to be true for all $0<r<r_-$, the condition~\eqref{c2} may still hold unless $F$ oscillates rapidly in the region $0<r<r_-$, in which case this would lead to a nonphysical solution.

\paragraph*{\textbf{4. Angular velocities.}} The angular velocity $\Om$ of the rotating hole~\eqref{r1} is\footnote{In Ref.~\cite{az3}, $\Om$ was unintentionally misprinted as $g_{\ta\phi}=\Om g_{\ta\ta}\sin^2\ta$. This is obviously a mistake since $g_{\ta\phi}\equiv 0$.} $\Om\equiv -g_{t\phi}/g_{\phi\phi}=2af/\Si$: This is the angular velocity, attributable to dragging effects, of freely falling particles initially at rest at spatial infinity as they reach the point ($r,\ta$). As $r\to\infty$, $\Om\to 2Jr^{-3}$, where $J=Ma$ is the angular momentum of the rotating hole. The angular velocity of the horizon $\Om_H$ is taken as $\Om(r_+)$. Using $\Si(r_+)=(r_{+}{}^2+a^2)^2$ along with~\eqref{sl2}, we obtain
\begin{equation}\label{sl3}
\Om_H=\frac{2af(r_+)}{\Si(r_+)}=\frac{ar_{+}{}^2[1-F(r_+)]}{(r_{+}{}^2+a^2)^2}=\frac{a}{r_{+}{}^2+a^2}.
\end{equation}

The four-velocity of the fluid elements is~\cite{az3,conf}
\begin{equation}
e^{\mu}_t=(r^2+a^2,0,0,a)/\sqrt{\ro^2 \De}.
\end{equation}
This can be written in terms of the timelike $t^{\mu}=(1,0,0,0)$ and spacelike $\phi^{\mu}=(0,0,0,1)$ Killing vectors as
\begin{equation}
e^{\mu}_t=N (t^{\mu}+\om \phi^{\mu}),
\end{equation}
with $N=(r^2+a^2)/\sqrt{\ro^2 \De}$ and $\om=a/(r^2+a^2)$ being the differentiable angular velocity of the fluid. Since the norm of the vector $t^{\mu}+\om \phi^{\mu}$, $1/N^2$, is positive only for $\De>0$, which corresponds to the region $r>r_+$, the fluid elements follow timelike world lines only for $r>r_+$. As $r\to r_+$, $\om$ approaches the limit $a/(r_+{}^2+a^2)$ that is the largest angular velocity of the fluid elements and this equals the angular velocity of the event horizon~\eqref{sl3}. So, the fluid elements are dragged with the angular velocity $\Om_H$ as all falling objects. At the event horizon, $t^{\mu}+\om \phi^{\mu}$ becomes null and tangent to the horizon's null generators.\\

\paragraph*{\textbf{5. Separability of Hamilton-Jacobi equation for neutral particles.}} With $S$ and $\tau$ denoting the Hamilton's principal function and proper time, the Hamilton-Jacobi equation takes the form
\begin{equation}\label{h1}
    2S_{,\tau}=g^{\mu\nu}S_{,\mu}S_{,\nu}\quad (S_{,\mu}\equiv \partial S/\partial\mu,\text{ etc}).
\end{equation}
With the elements of the inverse metric of~\eqref{r1} given by
\begin{align*}
&g^{tt}=\frac{\Si}{\ro^2\De},\,g^{t\phi}=\frac{2af}{\ro^2\De},\,g^{\phi\phi}=-\frac{\De-a^2\sin^2\ta}{\ro^2\De\sin^2\ta}\\
&g^{rr}=-\frac{\De}{\ro^2},\,g^{\ta\ta}=-\frac{1}{\ro^2},
\end{align*}
we expand the r.h.s. of~\eqref{h1} as\footnote{The coefficient of $(\partial S/\partial t)^2$ in Eq.~166 of Ref.~\cite{mtbh} should read $\Si/(\ro^2\De)$ instead of $\Si^2/(\ro^2\De)$.}
\begin{align}
2S_{,\tau}=&\frac{\Si}{\ro^2\De}(S_{,t})^2+\frac{4af}{\ro^2\De}S_{,t}S_{,\phi}-
\frac{\De-a^2\sin^2\ta}{\ro^2\De\sin^2\ta}(S_{,\phi})^2\nn\\
\quad &-\frac{\De}{\ro^2}(S_{,r})^2 -\frac{1}{\ro^2}(S_{,\ta})^2\\
\quad  =& \frac{[(r^2+a^2)S_{,t}+aS_{,\phi}]^2}{\ro^2\De}-\frac{[a\sin^2\ta\, S_{,t}+S_{,\phi}]^2}{\ro^2\sin^2\ta}\nn\\
\label{h2}\quad &-\frac{\De}{\ro^2}(S_{,r})^2 -\frac{1}{\ro^2}(S_{,\ta})^2.
\end{align}

For neutral particles, we assume, as usual,
\begin{equation}\label{h3}
    S=\frac{\ep}{2}\tau-Et+L\phi+S^r(r)+S^{\ta}(\ta)
\end{equation}
where $\ep=0$ for null geodesics and $\ep=1$ for timelike ones, and $L$ and $E$ are the conserved momentum and energy per unit mass of the particle (the mass is defined by $p_{\mu}p^{\mu}=m^2$, where $p^{\mu}$ is the 4-momentum of the particle). This ansatz allows us to bring Eq.~\eqref{h2} to the following form~\cite{mtbh}:
\begin{multline}\label{h4}
    \De (\dd S^{r}/\dd r)^2-\frac{[(r^2+a^2)E-aL]^2}{\De}+(L-aE)^2+\ep r^2\\
    (\dd S^{\ta}/\dd \ta)^2+(L^2\csc^2\ta-a^2E^2+\ep a^2)\cos^2\ta =0,
\end{multline}
which separates as
\begin{align*}
& \De (\dd S^{r}/\dd r)^2=\frac{[(r^2+a^2)E-aL]^2}{\De}-\mathcal{L}-(L-aE)^2-\ep r^2\\
& (\dd S^{\ta}/\dd \ta)^2=\mathcal{L}-(L^2\csc^2\ta-a^2E^2+\ep a^2)\cos^2\ta
\end{align*}
where $\mathcal{L}$ is a constant. This yields
\begin{equation}\label{h5}
    S=\frac{\ep}{2}\tau-Et+L\phi+\int^r\frac{\sqrt{\mathcal{R}(r)}}{\De}\dd r+\int^{\ta}\sqrt{\Theta(\ta)}\dd \ta
\end{equation}
where
\begin{align}
\label{h6}&\mathcal{R}(r)\equiv [(r^2+a^2)E-aL]^2-\De[\mathcal{L}+(L-aE)^2+\ep r^2]\\
\label{h7}&\Theta(\ta)\equiv \mathcal{L}-[L^2\csc^2\ta+a^2(\ep -E^2)]\cos^2\ta.
\end{align}
Note that the only dependence on $f$ in these last three equations is through $\De=r^2-2f+a^2$.

The basic equations are derived on setting to 0 the partial derivatives of $S$, as given by Eqs.~\eqref{h5}--\eqref{h7}, with respect to the constants ($\ep,\,L,\,E,\,\mathcal{L}$). Skipping the details of the calculations (similar derivations are done on p.~345 of Ref.~\cite{mtbh}), the basic equations of geodesic motion take the following forms where the dot denotes the derivative with respect to proper time $\tau$:
\begin{align}
&\ro^4\dot{r}^2=\mathcal{R},\,\ro^4\dot{\ta}^2=\Theta,\nn\\
\label{h8}&\ro^2\De\dot{\phi}=2aEf+(\ro^2-2f)L\csc^2\ta,\\
&\ro^2\De\dot{t}=E\Si-2aLf,\nn
\end{align}
where we have used\footnote{In Eq.~184 of Ref.~\cite{mtbh}, one should replace $\Si^2$ by $\Si$.} $\Si=\ro^2\De+2f(r^2+a^2)$ in the last equation. These basic equations are valid for any rotating regular or singular metric~\eqref{r1}. They generalizes the basic equations for metric derived in refs.~\cite{mtbh,circ}.

As an application of~\eqref{h8} (further applications are given in Ref.~\cite{prep}), we determine the condition(s) under which circular motion exists in the equatorial plane $\ta=\pi/2$. With $\ta=\pi/2$, the second equation in~\eqref{h8} [$\ro^4\dot{\ta}^2=\Theta$] reduces to $\mathcal{L}\equiv 0$; then, the first one yields
\begin{equation}\label{h9}
    r^2\dot{r}=\pm V^{1/2},
\end{equation}
where
\begin{multline}\label{h10}
V(r,\ep,a,L,E)\equiv \mathcal{R}(\mathcal{L}=0)\\
= [(r^2+a^2)E-aL]^2-\De[(L-aE)^2+\ep r^2].
\end{multline}

Circular orbits in the equatorial plane satisfy both conditions
\begin{equation}\label{h11}
    V=0\text{ and }V_{,r}=0,
\end{equation}
which can be solved for ($E,L$). The expression of $E^2$ reads
\begin{widetext}
\begin{equation}\label{h12}
E^2=\frac{8 (a^2-\Delta )^2 \Delta +2 r (a^2-\Delta ) \Delta  \De^{\prime}-a^2 r^2 \De^{\prime}{}^2\pm 2 \sqrt{2} a |\Delta|  \sqrt{(2
a^2-2 \Delta +r \De^{\prime})^3}}{r^2 [16 \Delta ^2+r^2 \De^{\prime}{}^2-8 \Delta  (2 a^2+r \De^{\prime})]}
\end{equation}
\end{widetext}
(here, $\De^{\prime}\equiv \dd \De/\dd r$), where we have assumed, without loss of generality, $a>0$. The upper sign in~\eqref{h12}, and in the following equations, corresponds to counterrotating particles, or retrograde circles, with $L<0$, and the lower sign to corotating particles, or direct circles, with $L>0$.

In order for $E^2$ to be real, the first obvious condition is
\begin{equation}\label{h12b}
    2a^2-2 \Delta +r \De^{\prime}\geq 0.
\end{equation}
Such a condition is never mentioned in the literature most likely because it is satisfied by astrophysical requirements demanding $r$ to be larger than the radius of the event horizon. The condition is static in that it does not depend on the rotation parameter $a$: With $\De=r^2-2f+a^2$, it reduces to
\begin{equation}\label{h13}
    2f-r f^{\prime}\geq 0.
\end{equation}
For the Kerr ($2f=2Mr$) and the Kerr-Newman ($2f=2Mr-Q^2$) solutions, \eqref{h12} reduces to $Mr\geq 0$ and $Mr\geq Q^2\equiv Mr_0$, respectively, and these constraints are satisfied in astrophysical applications. In the physical case $M>0$, the circle of radius $r_0\equiv Q^2/M$ is located inside the event horizon ($M+\sqrt{M^2-Q^2-a^2}$) of the Kerr-Newman \abh for all $a$, but it is outside the apparent horizon ($M-\sqrt{M^2-Q^2-a^2}$) for sufficiently small values of $a$. For $a^2>Q^2(M^2-Q^2)/M^2\equiv a_0{}^2$, $r_0$ is within the apparent horizon, too (it is obvious that $Q^2+a_0{}^2< M^2$ for $Q^2<M^2$).

Thus, in the physical case $M>0$, there is no circular equatorial motion for the Kerr \abh if $r<0$ and for the Kerr-Newman \abh if $r<Q^2/M$. For a rotating regular \aBH, as the rotating counterpart of the AGRSBH [given by~\eqref{r1} and~\eqref{s1}], or any rotating singular \aBH, clearly the constraint~\eqref{h12b} is satisfied within the event horizon $r_+$. That is, there exists a point $r_0<r_+$ such that~\eqref{h12b} is satisfied for $r\geq r_0$ where $r_0$ is a solution to $2f-r f^{\prime}= 0$. In fact, \eqref{h12b} is satisfied on the event horizon, since there $\De(r_+)\equiv 0$ and $\De^{\prime}(r_+)>0$ ($\De<0$ for $r<r_+$ and $\De>0$ for $r>r_+$); thus, it is also satisfied in the vicinity of the event horizon from within. Constraint~\eqref{h12b} is thus no harm for astrophysical applications.

The constraint~\eqref{h12b} is just a necessary condition for having circular equatorial motion. The requirement that $E^2>0$ imposes other physical constraints. Rewriting $E^2$ in the form
\begin{align}
\label{h14}&E^2=\frac{(\sqrt{x_1+x_2\sqrt{d}}\pm \sqrt{x_1-x_2\sqrt{d}}\,)^2}{r^2d}\\
\label{h14b}&E^2=\frac{2x_2}{r^2\sqrt{d}(\sqrt{x_1+x_2\sqrt{d}}\mp \sqrt{x_1-x_2\sqrt{d}}\,)^2}\\
&d\equiv 16 \Delta ^2+r^2 \De^{\prime}{}^2-8 \Delta  (2 a^2+r \De^{\prime})\nn\\
&2x_1\equiv 8 (a^2-\Delta )^2 \Delta +2 r (a^2-\Delta ) \Delta  \De^{\prime}-a^2 r^2 \De^{\prime}{}^2\nn\\
&2 x_2\equiv |2 \Delta ^2-2 a^2 \Delta -a^2 r \De^{\prime}|\nn
\end{align}
we see that $E^2>0$ if
\begin{equation}\label{h15}
    d>0\text{ and }x_1>0.
\end{equation}
Note that, since $x_1{}^2-x_2{}^2 d=8a^2\De^2(2a^2-2 \Delta +r \De^{\prime})^3$, the constraints $x_1>0$ and~\eqref{h12b} ensure the positiveness of $x_1-x_2\sqrt{d}$ in~\eqref{h14} and~\eqref{h14b}. The expressions~\eqref{h14} and~\eqref{h14b} of $E^2$ generalize Eq.~(2.12) of Ref.~\cite{circ} to all rotating regular or singular \BH.

Clearly, the two constraints~\eqref{h15} are not satisfied on the event horizon but they are manifestly satisfied outside of it for $\lim_{r\to\infty}\{d,x_1\}\to\{\infty,\infty\}$. They might be satisfied inside the apparent horizon, too.

From now on, we restrict ourselves to the astrophysical region that is located outside the event horizon. The limiting case $d=0$ results in a circular orbit for photons, for, in this case, the energy per unit mass generically diverges for retrograde circles [upper sign in~\eqref{h14}] as well as for direct circles [lower sign in~\eqref{h14b}]. Thus, the largest root $r_{\text{imb}}$ of
\begin{equation}\label{h16a}
   16 \Delta ^2+r^2 \De^{\prime}{}^2-8 \Delta  (2 a^2+r \De^{\prime})=0,
\end{equation}
after eliminating all common factors with $\sqrt{x_1+x_2\sqrt{d}}\pm \sqrt{x_1-x_2\sqrt{d}}$, provides the innermost circles for retrograde or direct circular motion. For the Kerr \BH, Eq.~\eqref{h14} reduces, after eliminating all common factors between numerator and denominator, to Eq.~(2.12) of Ref.~\cite{circ} and provides the innermost boundaries (imb's) of circular orbits for particles:
\begin{equation}\label{h16}
r_{\text{K}\,\text{imb}}=2M\{1+\cos[\tfrac{2}{3}\arccos(\pm a/M)]\}.
\end{equation}

\subsection{Special properties}

We specialize to the case of the AGRSBH where $F$ is given by~\eqref{s1}. Dropping the subscripts $\pm$, Eq.~\eqref{sl2} takes the form
\begin{equation}\label{r2}
    r^2-\frac{2Mr^4}{(r^2+Q^2)^{3/2}}+\frac{Q^2r^4}{(r^2+Q^2)^2}+a^2=0.
\end{equation}
As we noticed earlier, the locations of the horizons are functions of ($M,Q,a$) only. Unfortunately, one cannot solve~\eqref{r2} for $r$ in terms of ($M,Q,a$). For $Q^2/M^2\ll 1$, we obtain
\begin{align}
\label{r3}&r_{\pm}\simeq r_{\text{K}\;\pm}+c_{\pm}Q^2\\
&c_{\pm}=\frac{4 M\pm \sqrt{M^2-a^2}}{2 [a^2-M (M\pm \sqrt{M^2-a^2})]},\;c_+<0,\;c_->0\nn
\end{align}
where $r_{\text{K}\,\pm}=M\pm\sqrt{M^2-a^2}$ are the horizons of the Kerr \aBH. If $r_{\text{KN}\,\pm}$ denote the corresponding horizons of the Kerr-Newman hole
\begin{equation}\label{r4}
\hspace{-2.5mm}r_{\text{KN}\,\pm}=M\pm\sqrt{M^2-a^2-Q^2}\simeq r_{\text{K}\,\pm}\mp \frac{Q^2}{2\sqrt{M^2-a^2}},
\end{equation}
we obtain the order relations:
\begin{equation}\label{r5}
r_{\text{K}\,-}<r_{\text{KN}\,-}<r_-<r_+<r_{\text{KN}\,+}< r_{\text{K}\,+}.
\end{equation}
As far as the approximation, $Q^2/M^2\ll 1$ is valid, but this likely extends to all values of $Q^2$ within the limits of nonextremality; the horizons are ever closer than they are in Kerr or Kerr-Newman solutions.

It is also interesting to investigate the effects of nonlinear electrodynamics on the imb's of circular orbits for particles. For that purpose, we have developed enough tools in the previous section to tackle the question. We will do that in the approximation $Q^2/M^2\ll 1$ and compare the imb's for the Kerr, the Kerr-Newman, and the rotating regular counterpart of the AGRSBH.

Since we know the exact solutions for the imb's for the Kerr \abh [given by~\eqref{h16}], we do not need to look for and simplify any common factor(s) between the numerator and denominator of~\eqref{h14}. If $r_{\text{KN}\,\text{imb}}$ and $r_{\text{imb}}$ denote the imb's for the the Kerr-Newman \abh and the rotating regular counterpart of the AGRSBH, respectively, we look for solutions to~\eqref{h16a} of the forms
\begin{equation}\label{17h}
r_{\text{KN}\,\text{imb}}=r_{\text{K}\,\text{imb}}+C_{\text{KN}}Q^2,\,r_{\text{imb}}=r_{\text{K}\,\text{imb}}+CQ^2.
\end{equation}
Knowing the functions $f$ for the Kerr-Newman \abh and the rotating regular counterpart of the AGRSBH
\begin{align*}
&f_{\text{KN}}=M r-\frac{Q^2}{2} \\
&f=\frac{M r^4}{(r^2+Q^2)^{3/2}}-\frac{Q^2 r^4}{2 (r^2+Q^2)^2}
\end{align*}
we obtain the following expressions for $C_{\text{KN}}$ and $C$
\begin{align}
&C_{\text{KN}}=-\frac{4 a^2+4 r_{\text{K}\,\text{imb}} (r_{\text{K}\,\text{imb}}-3 M)}{3 r_{\text{K}\,\text{imb}} (r_{\text{K}\,\text{imb}}-M) (r_{\text{K}\,\text{imb}}-3 M)}\\
&C=-\frac{4 a^2+(3 M/2+17 r_{\text{K}\,\text{imb}}/2) (r_{\text{K}\,\text{imb}}-3 M)}{3 r_{\text{K}\,\text{imb}} (r_{\text{K}\,\text{imb}}-M) (r_{\text{K}\,\text{imb}}-3 M)}.
\end{align}
Now, Eq.~\eqref{h16} yields (1) $M<r_{\text{K}\,\text{imb}}<3M$ for retrograde circles resulting in $C>C_{\text{KN}}>0$ and (2) $3M<r_{\text{K}\,\text{imb}}<4M$ for direct circles resulting in $C<C_{\text{KN}}<0$. Thus, for a given value of $a^2$, the effects of nonlinear electrodynamics on the imb's is to (1) enlarge their size for retrograde circles and (2) reduce their size for direct circles:
\begin{align*}
&\text{retrograde motion: }&r_{\text{K}\,\text{imb}}<r_{\text{KN}\,\text{imb}}<r_{\text{imb}};\\
&\text{direct motion: }&r_{\text{K}\,\text{imb}}>r_{\text{KN}\,\text{imb}}>r_{\text{imb}}.
\end{align*}

The extremality condition and the common radius $r_{\text{ext}}$ of the merging horizons are solutions to~\eqref{r2} along with $\partial \De/\partial r=0$:
\begin{equation}\label{r6}
 1-\frac{M (r^2+4 Q^2) r^2}{(r^2+Q^2)^{5/2}}+\frac{2 Q^4 r^2}{(r^2+Q^2)^3}=0.
\end{equation}
A complete derivation of the extremality condition is provided in Appendix B. For $Q^2/M^2\ll 1$, the derivation yields
\begin{equation}\label{r7}
M^2\simeq a^2+4Q^2,\;r_{\text{ext}}\simeq M+\frac{3Q^2}{2M}.
\end{equation}
The same values for an extremal Kerr-Newman \abh are $M^2= a^2+Q^2$ and $r_{\text{KN}\,\text{ext}}= M$. The radius of the extremal rotating regular \abh is $3Q^2/(2M)$ larger than its Kerr-Newman counterpart.

For the same value of $M^2-a^2$, ones sees that a Kerr-Newman \abh may cumulate three more levels of the electric charge $(M^2-a^2)/4$ than a rotating regular one can do before the former becomes an extremal solution.

The latter conclusion extends to cases where the assumption $Q^2/M^2\ll 1$ is not valid, as Fig.~\ref{Fig1} (a) depicts. A consequence of that is to have no rotating and no static regular \bh for $Q^2<M^2$ but only regular non-black-hole solutions for all values of $a^2\geq 0$, as shown in Fig.~\ref{Fig1} (a). It is clear from that figure that a horizontal line $M^2/Q^2=C$ where $C>2.49$ intersects the extremality condition curve, of the rotating regular black hole~\eqref{r1} with $F$ given by~\eqref{s1}, at some critical value $a_c{}^2$ above which the rotating solution is no longer a \aBH. As $C$ gets closer to 2.49, $a_c{}^2$ approaches 0; if rotation increases a bit ($a^2\uparrow$), regular non-black-hole solutions become more favored than rotating \bh by nonlinear electrodynamics.

Another interesting conclusion driven from Fig.~\ref{Fig1} (a) is that the critical value $a_c{}^2$ for a rotating regular \abh is smaller than that for a rotating singular one. This may suggest the absence of superspinning regular holes.

\section{Conclusion\label{sec5}}

We have shown through examples from the literature that the final step in the NJA, which consists in bringing the generated rotating solution in EFCs to BLCs by real coordinate transformations, fails and that this is likely related to the complexification procedure. Since the latter procedure is, by itself, ambiguous, it seems there is no remedy to help overcome the situation.

In this work we have provided a method for generating rotating solutions in BLCs that is based partly on the NJA, but it avoids the complexification issues and employs physical arguments. The method applies equally to generate regular or singular rotating \abh and non-black-hole solutions as wormholes and so on~\cite{az3,conf}.

In this work, we have derived metric formulas in BLCs and in Kerr-like forms to generate generic rotating regular, as well as singular, \BH. These metrics are easy to handle, so we could provide simple treatments pertaining to the locations of the horizons and to the causality violations, could evaluate the SET and the scalar invariants, could provide the basic equation of geodesic motion for neutral particles, and could determine analytically the extremality condition.

We have concluded here and in Refs.~\cite{az3,conf} that the generic rotating \abh and non-black-hole solutions~\eqref{r1} are regular on the ring $\ro^2=0$, but physical entities are undefined there.

Another interesting conclusion, confirmed in Ref.~\cite{KB}, is that the rotating regular \aBH, the counterpart of the AGRSBH, has a much smaller electric charge and turns into a regular non-black-hole solution, for yet a small charge, well before its Kerr-Newman counterpart becomes naked singularity. This remark extends most likely to all rotating regular \bh that can be generated from known static regular ones. The nonlinear electromagnetic fields, due to the incursion of nonlinear electrodynamics in general relativity, are strong enough to help ``vanishing" the horizons, for still small charges, well before their Kerr-Newman counterparts can do so. Two other effects of nonlinear electrodynamics in general relativity are the absence of superspinning \bh and the convergence of the imb's of circular orbits for particles to the Kerr values having the same value of the rotation parameter.

We have reached the conclusion that causality violations do not occur in the region $0\leq r<r_-$, including small negative values of $r$ for all rotating regular \BH. By symmetry of the static regular \BH, this conclusion extends down to $-r_-$.

The still remaining open issues are the determination of the electromagnetic potential and energetics, as well as details of geodesic motion in the geometry, of a rotating regular \aBH.

\section*{\large Appendix A: Einstein equations}
\renewcommand{\theequation}{A.\arabic{equation}}
\setcounter{equation}{0}

The purpose of this section is to show that the general rotating solution~\eqref{r1} is solution to Einstein equations, $G_{\mu\nu}=T_{\mu\nu}$, where $T^{\mu\nu}$ is of the form~\eqref{rv1}. Consider, without specifying the form of the function $f(r)$,  the rotating solution~\eqref{r1}. For this solution, the basis $(e_t,\,e_r,\,e_{\ta},\,e_{\phi})$, dual to the 1-forms~\eqref{rv2}, reads
\begin{align}
&e^{\mu}_t=\frac{(r^2+a^2,0,0,a)}{\sqrt{\ro^2 \De}},\;e^{\mu}_r=\frac{\sqrt{\De}(0,1,0,0)}{\sqrt{\ro^2}},\;\nn\\
\label{B1}&e^{\mu}_{\ta}=\frac{(0,0,1,0)}{\sqrt{\ro^2}},\;e^{\mu}_{\phi}=-\frac{(a\sin^2\ta,0,0,1)}{\sqrt{\ro^2}\sin\ta},
\end{align}
and the nonvanishing components of the Einstein tensor $G_{\mu\nu}$ read
\begin{widetext}
\begin{align}
&G_{tt}=\frac{4 f^2+2 r [r^2+a^2 (2-\cos ^2\theta )] f^{\,\prime}-2 f [r^2+a^2 (2-\cos ^2\theta )+2 r f^{\,\prime}]-a^2
\sin ^2\theta  \rho ^2 f^{\,\prime\prime}}{\rho ^6},\nn\\
\label{B2}&G_{rr}=\frac{2 (f-r f^{\,\prime})}{\rho ^2 \Delta },\;G_{\theta \theta }=-\frac{2 (f-r f^{\,\prime})+\rho ^2 f^{\,\prime\prime}}{\rho ^2},\\
&G_{t\phi}=\frac{a \sin ^2\theta  [4 f (a^2+r^2+r f^{\,\prime})-4 f^2-(a^2+r^2) (4 r f^{\,\prime}-\rho ^6 f^{\,\prime\prime})]}{\rho
^6},\nn\\
&G_{\phi \phi }=\frac{\sin ^2\theta }{\rho ^6}\{4 a^2 \sin ^2\theta  f^2-f [2 (a^2+r^2) [r^2+a^2 (2-\cos ^2\theta
)]+4 a^2 r \sin ^2\theta  f^{\,\prime}]\nn\\
&\quad\quad\quad +(a^2+r^2) [2 r [r^2+a^2 (2-\cos ^2\theta )] f^{\,\prime}-(a^2+r^2)\rho ^2 f^{\,\prime\prime}]\} \nn.
\end{align}
\end{widetext}

If the solution~\eqref{r1} satisfies $G_{\mu\nu}=T_{\mu\nu}$, the components of the SET~\eqref{rv1} are expressed in terms of $G_{\mu\nu}$ as: $\ep=e^{\mu}_te^{\nu}_tG_{\mu\nu}$, $p_r=e^{\mu}_re^{\nu}_rG_{\mu\nu}=-g^{rr}G_{rr}$, $p_{\ta}=e^{\mu}_{\ta}e^{\nu}_{\ta}G_{\mu\nu}=-g^{\ta\ta}G_{\ta\ta}$, and $p_{\phi}=e^{\mu}_{\phi}e^{\nu}_{\phi}G_{\mu\nu}$. Using~\eqref{B1} and~\eqref{B2}, we arrive at~\eqref{11ab}.

\section*{\large Appendix B: Extremality condition}
\renewcommand{\theequation}{B.\arabic{equation}}
\setcounter{equation}{0}

We intend to find the extremality condition by solving both Eqs.~\eqref{r2} and~\eqref{r6}. Let
\begin{align}
\label{A1}&u^2\equiv a^2/Q^2,\;2s\equiv |Q|/M,\\
\label{A2}&x^2\equiv r_{\text{ext}}{}^2/Q^2,\;t=z^2\equiv x^2+1>1,
\end{align}
where the variables $x$ and $s$ have been used in Ref.~\cite{reg}. Equations~\eqref{r2} and~\eqref{r6} read, respectively,
\begin{align}
\label{A3}&\frac{z^3}{z^2-1}-\frac{1}{s}+\frac{1}{z}+\frac{u^2 z^3}{(z^2-1)^2}=0,\\
\label{A4}&1-\frac{1}{2 s} \frac{(z^2+3) (z^2-1)}{z^5}+\frac{2 (z^2-1)}{z^6}=0.
\end{align}
Solving~\eqref{A4} for $s$ and using the result in~\eqref{A3}, we arrive at
\begin{align}
\label{A5}&s=\frac{z (-3+2 z^2+z^4)}{2 (-2+2 z^2+z^6)},\\
\label{A6}&1-3 t-3 (u^2-2) t^2-(5+u^2) t^3+t^4=0.
\end{align}
Equation~\eqref{A6} admits one and only one real root greater than 1 for all $u^2\geq 0$: This is the root
\begin{equation}\label{A7}
t=\frac{5+u^2}{4}+\frac{\sqrt{W}}{2}+\frac{1}{2} \Big[Z+\frac{29+111 u^2+27 u^4+u^6}{4 \sqrt{W}}\Big]^{1/2},
\end{equation}
where
\begin{align}
&U=\sqrt{428+828 u^2+963 u^4+16740 u^6-1620 u^8-108 u^{10}},\nn\\
&V=(36+27 u^2+144 u^4-18 u^6+\sqrt{3} U)^{1/3},\nn\\
&W=2-u^2+3 (u^2-2)+\frac{1}{4} (5+u^2)^2+\frac{V}{18^{1/3}}\nn\\
\label{A8}&\quad +\frac{\left(\frac{2}{3}\right)^{1/3} (1-15 u^2+3 u^4)}{V},\\
&Z=u^2-2+3 (u^2-2)+\frac{1}{2} (5+u^2)^2-\frac{V}{18^{1/3}}\nn\\
&\quad -\frac{\left(\frac{2}{3}\right)^{1/3} (1-15 u^2+3 u^4)}{V}.\nn
\end{align}

With the expression of $t$ given by~\eqref{A7} and~\eqref{A8}, the extremality condition $M^2/Q^2=1/(2s)^2$ reads, substituting in~\eqref{A5},
\begin{equation}\label{A9}
\frac{M^2}{Q^2}=\frac{1}{t}\Big(\frac{t^3+2t-2}{t^2+2t-3}\Big)^2.
\end{equation}
Using~\eqref{A6} to eliminate all powers higher than 3, we arrive at
\begin{widetext}
\begin{equation}\label{A10}
\frac{M^2}{Q^2}=\frac{(84+97 u^2+21 u^4+u^6) t^3+3 (2+u^2) (-20+9 u^2+u^4) t^2+(56+38 u^2+3 u^4) t-(19+13
u^2+u^4)}{(37+17 u^2+u^4) t^3+3 (-21+7 u^2+u^4) t^2+(35+3 u^2) t-(9+u^2)}.
\end{equation}
\end{widetext}
The plot of Fig.~\ref{Fig1} (a) is a parametric plot of $1/(2s)^2$ versus $u^2$, and that of Fig.~\ref{Fig1} (b) is a parametric plot of $t-1$ versus $u^2$.

The limit $u^2\to0$ in~\eqref{A10} provides the extremality condition for the AGRSBH:
\begin{equation}\label{A11}
\text{for AGRSBH, }\frac{M^2}{Q^2}=\frac{216 t_s{}^2-112 t_s+65}{85 t_s{}^2-39 t_s+28},
\end{equation}
where the cubic terms have been eliminated using $t_s{}^3-4 t_s{}^2+2 t-1=0$, to which $t_s$ is the only real solution:
\begin{equation}
\hspace{-3.27mm}t_s=\frac{8+(332-12 \sqrt{321})^{1/3}+(332+12 \sqrt{321})^{1/3}}{6}.
\end{equation}
Numerically, the r.h.s. of~\eqref{A11} is $2.48641\simeq 2.49$, which is the value at which the plot of Fig.~\ref{Fig1} (a) intersects the vertical axis, and $t_s\simeq 3.51155$. The plot of Fig.~\ref{Fig1} (b) intersects the vertical axis at $t_s-1\simeq 2.51155$ yielding $r_{\text{ext}}/|Q|=\sqrt{t_s-1}\simeq 1.58$, as found in Ref.~\cite{reg}.


\end{document}